\documentstyle[12pt]{article}
\begin{document}

\begin{center}
{\bf PROBING SHORT RANGE NUCLEON CORRELATIONS IN  HIGH ENERGY HARD
QUASIELASTIC {\boldmath $pd$} REACTIONS \\ }
\vspace{4mm}
L.~Frankfurt$^{1,2}$, E.~Piasetsky$^{1}$, M.~Sargsyan$^{1,3}$,
M.~Strikman$^{4,5}$

\vspace{1cm}
\begin{center}
{\it
$^1$ Raymond and Beverly Sackler Faculty of Exact Science, \\
School of Physics and Astronomy, Tel Aviv University,\\
Tel Aviv, 69978, Israel\\
$^2$ On leave of absence from St. Petersburg Nuclear Physics Institute,
\\St. Petersburg, Russia\\
$^3$ Also at the Yerevan Physics Institute, Yerevan, 375036, Armenia\\
$^4$ Pennsylvania State University, University Park, PA, 16802, USA\\
$^5$ Also at the St. Petersburg Nuclear Physics Institute, St. Petersburg,
Russia}
\end{center}

\vspace{4mm}

\end{center}
\begin{abstract}
We show that the strong dependence of the amplitude for $NN$ hard scattering on
the collision energy can be used to magnify the effects of short range nucleon
correlations  in  quasielastic  $pd$  scattering.  Under  specific kinematical
conditions the  effect of initial and final state interactions can be accounted
for by rescaling the cross section calculated within  the   plane wave impulse
approximation. The feasibility to investigate the role of relativistic effects
in the deuteron wave function  is demonstrated  by comparing the predictions of
different  formalisms. Binding effects due to short range correlations in
deuteron are discussed as well.
\end{abstract}
\vspace{10mm}

\section{Introduction}
The deuteron is the simplest and best understood nuclear system bound by strong
interactions. As such it has received an extraordinary amount of attention.
The binding energy of the deuteron is rather small  compared to the nuclear
potential and to the binding energy of heavier nuclei and the average
distance between the nucleons in the deuteron is larger than the typical
distance  in nuclei. At first glance these facts would seem to discourage one
from using the deuteron for studying  short range correlations (SRC) between
nucleons. However, a conventional theory of the deuteron based on a
phenomenological potential of the nucleon-nucleon interaction with a
repulsive core at small distances predicts a fairly substantial contribution
of  high-momentum nucleon components in the wave function (WF) \cite{Brn}.
Note that the Fourier transform of the deuteron WF implies that high momentum
nucleon components originate predominantly from SRC.

The recent measurements of elastic  magnetic, electric and quadrupole
form factors at $Q^2~\leq~1~GeV^2$ \cite{emt} as well as previous data for
elastic \cite {Arnold75,Arnold87} and nearthreshold $ed$ \cite{Rock}
scattering at  $Q^2$ up to $10~GeV^2$, elastic $pd$ scattering \cite{pd:el}
and  the backward production of secondary $p$, $\pi$ on the
deuteron \cite{fb1,fb2} are all consistent with a conventional
nonrelativistic theory of the deuteron which has a  core in the NN  force
(see e.g. \cite{FS:rep81}). All these data show that the probability to
find   a nucleon with momentum above $0.3~GeV/c$ in  the deuteron wave
function is at the level of  $3-5\%$ and furthermore, that the $D$ wave
dominates for  the high momentum components.

Although the above experiments establish the presence of SRC in the deuteron
any detailed knowledge of the  high momentum  components in the deuteron WF
is quite limited. The elastic  and inclusive $(e,e')$ cross sections depend
on an  integral over  nucleon momenta in the deuteron. Also, the inclusive
character of the  backward particle production does not allow the unambiguous
reconstruction of the space-time picture of these processes.

  High  energy wide angle exclusive reactions belong to the special class of
processes which can probe rather directly the structure of SRC in the deuteron.
The ability to measure simultaneously  the spectator and the interacting
nucleons opens up a unique opportunity to  cross check various  interpretations
of SRC  dominated phenomena. This is the reason why we consider the high energy
exclusive reactions to be the most straightforward method for investigating
SRC. It is a timely venture to investigate theoretically these processes on
the deuteron since the high accuracy facilities for studying these processes
experimentally with  high intensity proton \cite{Eva} and  electron
\cite{Cebaf} beams will be available very soon.

In this paper we analyze the capacity for studying SRC in the deuteron by a
complete kinematical measurement of the hard proton-deuteron  quasielastic
scattering. The salient feature of this reaction (discussed in section 2) is
that within the framework of the plane wave impulse approximation (PWIA) the
strong energy dependence of the hard $pp$ elastic cross section boosts the
contribution of the high momentum components of the  wave function\cite{FFS}.
We show (section 3) that initial and final state interactions do not spoil
this conclusion. In section 4 we show that there is a significant difference
between two varying  approaches to the description of deuteron structure
which  enables us to demonstrate the importance of relativistic effects.
The fact that the  processes are sensitive to high momentum components  of
the deuteron WF can be also used to investigate models of the  EMC effect.
In order to demonstrate that,  we calculate the effect of the suppression of
point-like configurations due to SRC \cite{FS85}. In the Appendix we
present some details of the calculation of the initial and final state
interactions.

\section{Plane Wave Impulse Approximation}

When calculating the  cross sections of hard processes it is necessary to
include relativistic kinematics  and the fact that high energy processes are
developed along the light cone (LC). We calculate the cross section for the
quasielastic $d(p,2p)n$ reaction within the light-cone  impulse approximation.
In Section 4 we show that for small spectator momenta in the deuteron the
predictions based on LC mechanics  are indistinguishable from  those based on
a nonrelativistic theory. Within the LC impulse approximation the cross
section is a convolution of the elementary $pp$ cross section and the deuteron
light cone density function \cite {FS:rep81}:

\begin{eqnarray}
\frac{d^{6}\sigma}{(d^{3}p_{3}/E_{3})(d^{3}p_{4}/E_{4}) }
= \kappa_{p}{1\over 2\pi}{\tilde s^{2}-4m^{2}\tilde s\over 2m\cdot|\vec p_1|}
\cdot
\frac{d\sigma}{dt}^{pp}(\tilde s,t)\cdot {\rho_{D}(\alpha,p_{t}^{2})\over
\alpha^2 p_{D-}}\delta(p_{s+} - {m^{2}+p_{t}^{2} \over m(2-\alpha)})
\label{x_lc}
\end{eqnarray}

where
\begin{eqnarray}
\alpha & = & \alpha_{4} + \alpha_{3} - \alpha_{1} \ ; \
\alpha_{i} = 2{p_{i-}\over p_{D-}} \equiv  2{E_{i}-p_{i}^z \over
E_{D}-p_{D}^z} \ ; \ p_{t}  = p_{3}^t + p_{4}^t
\nonumber \\
\nonumber \\
\tilde s & = & ({p_3}+{p_4})^{2} \approx 2m^2 + 2E_1m\alpha - {\alpha \over 2}
\left (4{m^2+p^2_t \over \alpha(2-\alpha)} - m_{D}^2 \right )
 \ ; \ t = ({\it p_1}-{\it p_3})^{2}
\label{alpha}
\end{eqnarray}
where ${ p_{D}} =(E_{D},\vec{p_{D}})$, ${p_{1}}=(E_{1},\vec{p_{1}})$,
${ p_{3}}=(E_{3},\vec{p_{3}})$, ${p_{4}} =(E_{4},\vec{p_{4}})$,
${p_{s}} =(E_{s},\vec{p_{s}})$ are the four - momenta of the target nucleus,
incoming, scattered, produced and spectator nucleons, respectively
(see e.g. Fig.4c). The indices $"t"$ and $"z"$ denote the transverse and
longitudinal direction with respect to the incoming proton momentum
$\vec p_{1}$. The $"+"$ and $"-"$ denote the energy and longitudinal
components of the four-momenta in the light cone reference frame:
$p_{\pm}~=~E~\pm~p_z$. The factor $\kappa_{p}$ accounts for the effects
of initial and final state interactions.
In the plane wave approximation $\kappa_p = 1$.

The light-cone density function of deuteron expressed in terms of $S$ and $D$
components of the wave function is  as follows \cite {FS79}:

\begin{eqnarray}
\rho_{D}(\alpha,p_{t}^{2}) & = & {u(k)^{2} + w(k)^{2}) \over 2 - \alpha} \cdot
\sqrt{m^{2} + k^{2}} ; \nonumber \\
k & \equiv & k(\alpha,p_t)  = \sqrt{{m^{2}+p_{t}^{2} \over
\alpha(2-\alpha)} - m^{2}} \ ; \ (0 < \alpha < 2)\label{rho}
\end{eqnarray}

Note that in  eq.~(\ref{rho}) we ignore  possible admixtures of  non-nucleonic
degrees of freedom  in the WF. This is justified  by the analysis of
high energy processes on the deuteron which indicate that the nucleon-nucleon
component of deuteron WF dominates in the region of momenta below
$(0.5~\sim~0.6)~GeV/c$  (for a review see \cite {FS:rep88}).

We write $ \frac{d\sigma}{dt}^{pp}$ for the invariant cross section for
elastic $pp$ scattering. We use phenomenological $pp$ differential cross
section  parameterized by:

\begin{eqnarray}
 \frac{d\sigma}{dt}^{pp}(s,t) & = & 45.0 {\mu b\over sr GeV^{2}}\cdot
\left ({10\over s} \right )^{10}\cdot \left ( {2\cdot t \over 4m^{2}-s}\right
)^{-4\gamma}\cdot\nonumber \\
&   &  \nonumber \\
&   & \left [ 1 + \rho_{1}\sqrt{s\over GeV^{2}}\cdot\cos{\phi(s)} +
{\rho_{1}^{2}\over 4}{s\over GeV^{2}} \right ]\cdot F(s,\theta_{c.m.})
\label{x_pp}
\end{eqnarray}

where $\rho_{1} = 0.08$, $\gamma = 1.6$ and $\phi(s) = {\pi\over 0.06}
\ln(\ln[s/(0.01 GeV^{2})])^{-2}$.   At $s~>~\sim~6.5 \ GeV^{2}$
eq~.(\ref{x_pp}) reproduces the parameterization of Ralston and Pire
\cite {RLPR} for $\theta_{c.m.}~=~90^{0}$ while at fixed $s$ it reproduces
the parameterization of Brodsky \cite {BR:pr}.
The function $F(s,\theta_{c.m.})$   is  used to  further adjust the
phenomenologically motivated parameterization of the experimental
data in the range $60^{0}~\leq~\theta_{c.m.}~\leq~90^{0}$.
We present the function $F(s,\theta_{c.m.})$ for four typical $c.m.$ scattering
angles (Fig.1). At the limit of large $s$ and large $c.m.$ scattering angle the
phenomenological motivated approximation is the best and  $F(s,\theta_{c.m.})$
approach unity. As far as we get from the ideal case the larger
$F(s,\theta_{c.m.})$ is.
For  cross sections at lower $s$ we use the phase-shifts  generated by the
{SAID} code \cite {ARNDT}.

We draw attention to the strong dependence on $s$ of the elementary cross
section. According to eq.~(\ref{x_pp}), for a fixed value of $-t$
($\geq 2 GeV^2$), the  $pp$ cross section behaves as $\sim s^{-(10-4\gamma)}$
which according to eq.~(\ref{alpha}) will introduce a $\sim
\alpha^{-(10-4\gamma)}$ dependence in the PWIA cross section
(see eq.~(\ref{x_lc})). Thus, in  PWIA the cross section for
proton wide angle scattering on the deuteron is sensitive to the low $\alpha$
values of the  wave function \cite {FFS} i.e. to that high Fermi momentum
of the struck nucleon which is predominantly  parallel to the incoming beam.
This is the key feature that makes the process a useful tool to magnify SRC
effects. Indeed it complements other probes such as electrons which are
usually  more sensitive to the $\alpha \ge 1$ region.

We calculated the cross sections using eqs.(\ref{x_lc})-(\ref{x_pp}). They are
shown in Fig.2 as a function of the momentum ($p_s$) and polar
angle ($\theta_s$) of the spectator neutron with respect to the incoming
proton. One can see a significant enhancement of the cross section
for large momenta of the spectator nucleon  at  large projectile  momenta
($\geq 12 GeV/c$) and when the target proton momentum
($\theta_2 = \pi - \theta_s$) tends to be parallel to the projectile.
Thus the discussed  process is rather sensitive to the high momentum part of
the deuteron wave function which in turn is the result of the strong $s$
dependence of the elementary $pp$ cross section.
Because of this strong $s$ dependence it is necessary to pay special
attention to off-shell effects, which are proportional to  the difference
between the invariant energies at  the intermediate ($s_{in}$) and final
($\tilde s$) states. Since in LC mechanics only the  "-" and "t"  components
of momenta are conserved in an intermediate state, this difference is:
\begin{equation}
s_{in} - \tilde s = m_D\cdot (2E_s-m_D) + p_{1-}(p_2 - (p_D-p_s))_+ =
 m_D\cdot (2E_s-m_D) + {p_{1-}\over m\alpha}\cdot m_D(2E_s-m_D)
\label{off_sh}
\end{equation}
where $p_{2+} = {m^2+p_{st}^2\over m\alpha}$. The difference between
$s_{in}$ and $\tilde s$  does not increase with incoming energy since
$p_{1-}~=E_1-p_1~\rightarrow~0$. We estimate the size of the effect by
calculating  the cross section with eq.(\ref{x_lc}) for two values of the
invariant energy $\tilde s$ and $s_{in}$.
The two calculations were done for  the kinematics where the projectile is
parallel to the struck proton and the results are shown  in Fig.3.
Note  that the off-shell effects depend  only weakly on  the direction of
the struck nucleon momentum (see eq.(\ref{off_sh})). We see also that  the
off-shell effects cause at most $\approx 10\%$ change in the cross section
values for  spectator momenta up to $\sim~0.3~GeV/c$.

Since the  kinematics of quasielastic scattering on a deuteron is fully
determined by  the momenta of the two ejected protons  one can reconstruct
the geometry of the elementary $pp$ hard scattering. We calculate the cross
sections for three extreme geometries where the momentum of  the struck
nucleon is parallel, antiparallel and perpendicular to the direction of the
incident proton and  we refer to them accordingly as parallel, antiparallel
and perpendicular.


\section{The Effects of Initial and Final State Interactions}
 In order to extract usefull information from the data one has to account
for the initial  state interaction (ISI) of the incoming proton and  the  final
state interactions (FSI) between the outgoing protons and  the  spectator
neutron.

 We will choose   kinematical conditions such that the ISI and the FSI
corrections will be  small or, at least, will not introduce any additional
energy dependence in the spectra of scattered nucleons.

 We will use the eikonal approximation to  calculate the sum of the diagrams
shown in Fig.4 which describe the hard PWIA scattering (Fig.4a) and the lowest
order soft $NN$ rescatterings (Fig.4b-d).
The soft interactions between the  incident  and target protons as well as the
interactions between the  scattered and produced protons are included  in the
phenomenological parameterization of the $pp$ hard scattering amplitude.

Since at high energies the soft rescattering amplitude depends almost
linearly on the  energy of the collision $s$, the contribution of the ISI
diagram (Fig.4b) is equal to the  contribution of any one of the FSI
(Fig.4c,d) diagrams. Therefore, just one of the three diagrams (b,c,d) has to
be evaluated and the final result is given by the PWIA diagram plus three
times the contribution  from any one of the rescattering  diagrams
(see details in the Appendix). Since the soft scattering decreases
exponentially with transfered momentum  while the hard scattering has a
power law dependence it is convenient to factorize  the effect of
ISI and FSI into a scale factor $\kappa_p$:

\begin{eqnarray}
\kappa_{p}  &  =  & 1 -
{3 \over 2} \int {u(k)u(k') + w(k)w(k')[{3\over 2}{(kk')^{2}
\over k^2 k'^2} - {1\over 2}] \over u^{2}(k) + w^{2}(k) }
\left ( {\sqrt{m^2+k'^2} \over \sqrt{m^2+ k^2}} \right )^{{1\over 2}}
\cdot f(q_{t}) \cdot{d^2q_{t} \over (2\pi)^2} \ \ \ +  \nonumber \\
\nonumber \\
& &
{9 \over 16} \int {\left ( u(k_1)u(k_2) + w(k_1)w(k_2)[{3\over 2}{(k_1k_2)^{2}
\over k_1^2 k_2^2} - {1\over 2}] \right )
\left ( \sqrt{(m^2+ k_1^2)(m^2+ k_2^2)}\right )^{{1\over 2}}
\over \left ( u^{2}(k) + w^{2}(k) \right )\sqrt{m^2+k^2}}\cdot \nonumber \\
& & \ \ \ \ \ \ \ \ \ \ \ \ \ \ \ \ \ \ \ \ \ \ \ \ \ \
\ \ \ \ \ \ \ \
\ \ \ \ \ \ \ \ \ \ \ \ \ \ \ \ \ \ \ \ \ \ \ \ \
 \cdot f(q_{1t})f(q_{2t}){d^2q_{1t} \over (2\pi)^2} {d^2q_{2t} \over (2\pi)^2}
\label{kappa}
\end{eqnarray}
where $k=k(\alpha,\vec p_t)$, $k'=k'(\alpha,\vec p_t+ \vec q_{t})$,
$k_1 = k_1(\alpha,\vec p_t+ \vec q_{1t})$, $k_2 = k_2(\alpha,\vec p_t+
\vec q_{2t})$ are given by eq.~(\ref{rho}) and $f(q_{t})$  is the imaginary
part of the elastic $NN$ soft scattering amplitude. For simplification we
assumed  that $f(q_{t})~=~\sigma_{tot}^{NN}\exp{(-bq_{t}^2)}$ and neglected
nucleon spin rotations in the LC which may add uncertainties of the order
of 1\% (see  \cite{FS83} for  estimates).

The Eq.(\ref{kappa}) contains the distinctive feature of soft high energy
processes that the $\alpha$ component of the target nucleon momentum is
conserved in soft rescatterings. In Fig.5 we present the value of $\kappa_{p}$
as a function of $\alpha$ for different values of the spectator transverse
momenta ($p_s^t$). It shows that the ISI and FSI are significant for
$|\alpha-1|~\geq~\sim~0.3$ for any $p_s^t$ and at large $p_s^t$
($~>~\sim~0.1~GeV/c$), for a large range of $\alpha$. In these kinematical
conditions the contribution from higher order rescatterings is not negligible
and consequently our results could not be trustworthy. The important
feature of ISI and FSI is that at small $p_s^t$ the $\kappa_{p}$ is a  smooth
function of $\alpha$ for $0.7~\leq~\alpha~\leq~1.3$, which corresponds to
spectator momenta $|\vec p_s|~\leq~0.25~GeV/c$ in parallel geometry and $|\vec
p_s|~\leq~0.4~GeV/c$ in antiparallel geometry. In these cases the difference of
$\kappa$ from $1$ is typically $\leq~20\sim~25\%$  and the effects of initial
and final state interactions can be accounted for by rescaling the PWIA cross
section. Thus, quasielastic processes with small $p_s^t$ and $\alpha$  within
the above discussed kinematical range are well suited for  investigating high
momentum nucleon component in the deuteron.
It  means that under these conditions the deuteron is a good testing ground
for the studying the basic ingredients of the nucleon - nucleon interaction.

Note that in this paper we do not consider any  possible effects due to
color transparency  (see e.g. \cite{FS91} and  references therein) which could
suppress  considerably the effects of ISI and FSI especially at larger
projectile energies.
We will elaborate on the physics of color transparency in forthcoming
publications.


\section{Comparision Between  Different Approaches}

An interesting question is to what extent the effects due to the nucleon
relativistic motion in the deuteron are important and whether they can be
investigated experimentally. In  Fig.6 we show two time-ordered diagrams which
represent the standard impulse approximation contribution (A) with on - energy
shell $NN$ amplitude and the  relativistic contribution (B).
To  examine to what extent the relativistic contribution (diagram B) is
important we compare the results obtained in the light-cone  impulse
approximation with those of the  virtual nucleon formalism (VN) (see e.g.
\cite {West}):
\begin{eqnarray}
\frac{d^{6}\sigma}{(d^{3}p_{3}/E_{3})(d^{3}p_{4}/E_{4}) }
= \kappa_{p}{1\over 2\pi}{\tilde s^{2}-4m^{2}\tilde s\over 2m\cdot |\vec p_1 |}
\frac{d\sigma}{dt}^{pp}\cdot {u^{2}(p_s)+w^{2}(p_s) \over N(E)}\delta(E_{s}
 - (M_{D}-E)) \nonumber \\
\label{x_vn}
\end{eqnarray}
where $ E = E_3 + E_4 - E_1 $. The factor $N(E) = 2E/M_{D}$  accounts for
the conservation of baryon charge \cite{nm}.

In the LC impulse approximation we know  (see e.g. \cite{KOSP},
\cite{FS:rep81})
that the contribution of diagram B is negligible or included into the
definition of the deuteron wave function. By comparing the LC impulse
aproximation  to the VN approximation, where diagram B is included in
different way, we actually can learn about the  importance of diagram B.

It is interesting to note that we account for  ISI and FSI
in  the virtual nucleon impulse approximation by an
equation which is very similar to eq.(\ref{kappa}).
The only major difference with the  LC approximation is the different
argument (momentum) of the deuteron wave function.

The ratio of the cross sections  calculated in the LC impulse approximation
(eq.~(\ref{x_lc})) and in the virtual nucleon approach (eq.~(\ref{x_vn}))
is shown in Fig.7. The ratio is displayed as a function of the spectator
(neutron) momentum for the above defined parallel (antiparallel) and
perpendicular geometries. We use  deuteron wave functions calculated with the
Bonn\cite{Bonn} and Paris \cite{Paris} potentials. Formally,  the main
difference between the LC  and virtual nucleon formalisms is in the value of
the momentum at which the deuteron wave function is calculated by
eq.~(\ref{rho}). For the perpendicular geometry as well as for small
spectator momenta $\leq~0.1~GeV/c$  the 'internal' momentum $k$ and the
spectator momentum in the laboratory  system are  rather close to each other.
Under those conditions there is no substantial difference between the
relativistic and non-relativistic calculations. However, there is a
considerable ($\sim~20-25\%$) difference between these approaches even at
relatively low spectator momenta  of about $\sim 0.25~GeV/c$ where the
deuteron wave function is known with an accuracy of a few percent and where
the ISI and FSI can be replaced by a  rescaling factor (see previous section).

Another way to confront various approaches that account for the relativistic
motion of nucleons in the deuteron is to measure the asymmetry between the
cross sections for parallel and antiparallel geometries at light-cone
momentum fractions $\alpha$ and $2-\alpha$. We define the asymmetry parameter
$A(\delta)$ as \cite{FMS}:
\begin{equation}
A(\delta) = {f(1-\delta)-f(1+\delta) \over [f(1+\delta)+f(1-\delta)]/2}
\label{asm}
\end{equation}
where $f~=~{\alpha^2\sigma^{pD}\over (s^2-4m^2s)\cdot \sigma^{pp}}$
(see eq.~(\ref{x_lc})) and $\alpha~=~1~-~\delta$, $2~-~\alpha~=~1~+~\delta$.

Since the light-cone formulae (eq.~(\ref{x_lc}) and eq.~(\ref{kappa})) are
invariant under  exchange  of $\alpha$ and $2~-\alpha$  we get that
$A(\delta)~=0$. This is because the two nucleons in the deuteron are treated
symmetrically in light-cone quantum mechanics. In contrast, in the virtual
nucleon description  the interacting nucleon is  off-shell while the spectator
is on the mass shell. As a result the symmetry with respect to the
$\alpha~\Leftrightarrow~2-\alpha$ transposition is lost. The significance of
this effect can be seen in  Fig.8 which shows the asymmetry calculated in the
VN approximation according to eq.~(\ref{asm}). A significant difference
between the LC prediction ($A(\delta)~=~0$) and the VN one
$A(\delta)~\approx~0.5$ reveals itself at comparatively low spectator momenta
$p_{sz} \sim~-0.25~GeV/c$ for parallel and $p_{sz} \sim~0.34~GeV/c$ for
antiparallel geometries. The initial and final state interactions plays
insignificant role at low spectator momenta (cf. above discussion)  and
will even enhance the difference  at higher momenta.

One can also investigate the difference between the two approaches by  looking
at the symmetry of the cross sections under the transformation which changes
the sign of the spectator momenta $\vec p_s$. Such a symmetry is expected in
the VN approximation eq.(\ref{x_vn})  but not in the light-cone description
eq.(\ref{x_lc}).

It is worth mentioning that at the large energies of the experiment where one
measures the two emerging protons will yield a better resolution in $\alpha$
than in $p_s$ \cite{Yuan}. Therefore, from a practical point of view there is
an advantage  in testing the asymmetry in $\alpha$ rather than in $p_s$.
The ISI and FSI effects suppress the sensitivity to the various deuteron wave
functions as can be seen  in Figs. 7 and 8. The suppression occurs because
the integrals in eq.~(\ref{kappa}) are sensitive mainly to the lower momenta
in the deuteron wave function.

The fact that the  target nucleon is bound in a deuteron can  significantly
alter the picture of high-energy hard scattering especially for the  large
 momenta. One consequence of nucleon binding is the  suppression of
point - like configurations for  bound nucleons \cite {FS85}.  At  distances
where the two-nucleon interaction is dominated by the attractive part of the
nucleon-nucleon potential the probability of small size quark-gluon
configurations ( point-like configurations (PLC)) is suppressed  (color
screening phenomenon \cite {FS91}). For the discussion how this effect reveals
itself in the interaction of two nucleons see  \cite{KAF}. The amount of
suppression can be estimated by multiplying the deuteron LC wave function
by the factor \cite{FS85}:

\begin{equation}
\delta (k,t) = \left( 1 + \Theta (t_{0}-t)\cdot (1- {t_{0}\over t})\cdot
{{k^{2}\over m_{p}} +2\epsilon_{D} \over \Delta E} \right )^{-2}
\label{emc}
\end{equation}
where $\epsilon_{D}$  is the deuteron binding energy and $\Delta E$
($ \approx 0.6 \ GeV$) is a parameter  which characterizes  the  bound nucleon
excitation in the deuteron. The additional $t$ dependence ($ 1~-~{t_{0}\over
t})$ \cite {ptf} accounts for  the fact that  point - like configurations
dominate in the bound nucleon wave function for  reactions with  sufficiently
large transfered momentum $t$. We used the value $t_o~=~-2 \ GeV^2$.
In  Fig.9  we show the suppression  of PLC as a function of the spectator
momentum for the parallel, perpendicular and antiparallel geometries.
Except for very large spectator momenta  the effect is the same for the various
geometries. The  suppression  is relatively small in the parallel geometry
for $p_s~\geq~0.5~GeV/c$ due to the small energy transfer in this kinematical
condition. This is a consequence of the $t$ dependence in eq.~(\ref{asm}).
The  PLC suppression  decreases the cross section and  increases the difference
between the  LC and VN approximations for  parallel kinematics
(solid line in Fig.9). Note also that for $|t|~\gg~2~GeV^2$ the $t$ dependence
becomes weaker and that the symmetry under
$\alpha~\Leftrightarrow~2-\alpha$  transformation is restored.

\section{Conclusion}
 We summarize our results in Fig.10 where we display the cross section for hard
quasi-elastic proton scattering on the deuteron by taking into account the
various SRC effects discussed in this paper. We see  that ISI and FSI and
nuclear binding effects do not mask the predicted enhancement of high momentum
components for parallel kinematics.

We reiterate that a fully  kinematical measurement at the appropriate
kinematical conditions makes it possible to separate the  various effects we
discussed. For example, at neutron spectator momenta of about $250~MeV/C$,
where the deuteron wave functions are well known, any difference between a
bound  and  a  free  nucleon in parallel kinematics tends to increase the
differences between the light-cone and virtual nucleon results. Also the
contributions from ISI and FSI are small and have a relatively weak momentum
dependence.

We have shown that there is considerable difference for the asymmetry of cross
sections in parallel and antiparallel geometries when calculated in the LC and
VN approximations  in a kinematical region where ISI and FSI effects and
effects  of EMC are well understood. These differences allow us to
determine the importance of the vacuum diagram contribution.
Eventually, after we achieve some understanding of hard scattering, nucleon
binding, ISI and FSI effects, any new data will be able to teach us something
about the deuteron WF at momenta above the region studied thus far.
\vspace{5mm}

{\bf Acknowledgment} We wish to thank Jonas Alster, Steve Hepelmann,
Yael Mardor and Israel Mardor for helpful discussions. This work was supported
in part by the Basic Research Foundation administrated
by the Israel Academy of Science and Humanities,
by the U.S.A. - Israel Binational Science Foundation  Grant No. 9200126
and by the U.S. Department of Energy under Contract No. DE-FG02-93ER40771.

\section*{Appendix}
\begin{appendix}
To acount for  the intial state interaction of the incoming proton and of the
final state interaction of the scattered and knocked out protons  with  the
spectator neutron we need to calculate  the sum of the diagrams in Fig.4:
\begin{eqnarray}
T_D^{if} & = & T^h(q)\cdot \psi(\alpha,p_t)   \nonumber \\
%
& &- \int_{(b)} {T^h(s,t')\cdot T^s(k) \over ( (p_s-k)^2-m^2+i\varepsilon )
( (p_1-k)^2-m^2+i\varepsilon )}\cdot \psi(\alpha+\alpha_k,p_t+k_t)\cdot
{d^4k\over i(2\pi)^4}   \nonumber \\
& &\nonumber \\
& &- \int_{(c)} {T^h(s',t')\cdot T^s(k) \over ((p_s-k)^2-m^2+i\varepsilon )
((p_3+k)^2-m^2+i\varepsilon )}\cdot \psi(\alpha+\alpha_k,p_t+k_t)\cdot
{d^4k\over i(2\pi)^4}   \nonumber \\
& &\nonumber \\
& &- \int_{(d)} {T^h(s',t)\cdot T^s(k) \over ((p_s-k)^2-m^2+i\varepsilon )
((p_4+k)^2-m^2+i\varepsilon )}\cdot \psi(\alpha+\alpha_k,p_t+k_t)\cdot
{d^4k\over i(2\pi)^4} \nonumber \\
\label{a1}
\end{eqnarray}
where $s'~=~s~-~2(p_3+p_4)_t~k_t~-~k_t^2$, $t'~=~t~-~2p_{3t}k_t~-~k_t^2$ and
$\alpha_k~=~{k_-\over m}~\equiv~{k_0-k_3\over m}$.  $T^h$  is the amplitude
for the hard $pp$ elastic scatterings,  $T^s$  is the amplitude for
soft $pn$ rescatterings and $\psi(\alpha,p_t)$  is the deuteron
light cone wave function. Spin effects are not included.  The labels
"b", "c" and "d"  correspond to the  diagrams in Fig.4. It is well known that
the sum of these diagrams leads to lowest order formulae of the eikonal
approximation where the intermediate particles are on the mass shell.
We  calculate  only the diagram "b" since the calculation for the
other two rescattering diagrams is practicaly identical. In terms  of light
cone variables  $k~(k_+,k_-,k_t)$  the invariant phase volume is
$d^4k~\rightarrow~{1\over 2} dk_+dk_-dk_t$ and the denominator of integrand
"b" is given by:
\begin{eqnarray}
((p_s-k)^2-m^2+i\varepsilon)((p_1-k)^2-m^2+i\varepsilon) =
\ \ \ \ \ \ \ \ \ \ \ \ \ \ \ \ \ \ \ \ \ & \nonumber \\
(p_1-k)_+\left[ p_{1-} - k_- - {m^2+(p_1-k)_t^2\over p_{1+}-k_+}
+ i{\varepsilon\over p_{1+}-k_+} \right ]\cdot
&\nonumber \\
(p_s-k)_-\left[ p_{s+} - k_+ - {m^2+(p_1-k)_t^2\over p_{s-}-k_-}
+ i{\varepsilon\over p_{s-}-k_-} \right ]&
\label{a2}
\end{eqnarray}
To evaluate the integral we take residues over $dk_+$ and $dk_-$ which
correspond to on-shell   intermediate nucleons (namely the incident proton and
spectator neutron). Finally, for diagram "b" we obtain:
\begin{eqnarray}
"b" = {i\over 4} \int {T^h(s,t')\cdot T^s(k) \over s_1}\cdot
\psi(\alpha+\alpha_k,p_t+k_t)\cdot {d^2k_t\over (2\pi)^2}
\mid_{k_- = p_{1-} - {m^2+(p_s-k)_t^2\over p_{1+}-k_+} }
\label{a3}
\end{eqnarray}
where $s_1~=~(p_1-k)_+(p_s-k)_-~\approx~p_{1+}p_{s-}$ is the center of mass
energy for the incident proton and the spectator neutron rescattering.
As follows from eq.~(\ref{a3}) at sufficiently high energies and soft
rescatterings $k_-~\approx~0$ and $\alpha'~\approx~\alpha$ and therefore
the longitudinal ($\alpha$) component of wave function is not  changed by the
soft rescattering.

Similarly, for the diagrams "d" and "c" one gets the following expressions:
\begin{eqnarray}
"c" = {i\over 4} \int {T^h(s',t')\cdot T^s(k) \over s_3}\cdot
\psi(\alpha+\alpha_k,p_t+k_t)\cdot {d^2k_t\over (2\pi)^2}
\nonumber \\
\nonumber \\
"d" = {i\over 4} \int {T^h(s',t)\cdot T^s(k) \over s_4}\cdot
\psi(\alpha+\alpha_k,p_t+k_t)\cdot {d^2k_t\over (2\pi)^2}
\label{a4}
\end{eqnarray}
where the $s_3~\approx~p_{3+}p_{s-}$ and $s_4~\approx~p_{4+}p_{s-}$  are the
invariant energies corresponding to rescattering of the scattered and produced
protons with the spectator neutron.

For soft rescattering  $T^s(k)/s$ is practically independent to $s$. Since the
dependence  on  $t$  of the hard scattering is slower than the exponential
dependence of soft rescattering, one can neglect the difference between $t$
and $t'$ in the hard scattering amplitude. As a result we find that  the
diagrams of "b", "c" and "d" contribute equally. Therefore, the scattering
amplitude can be expressed  by the  sum of the Born diagram   plus three times
the contribution  of any one of the diagrams "b", "c", "d".
Factorizing the hard scattering amplitude from this sum,
including the spin dependence of the deuteron wave function and taking  the
square of the amplitude we obtain the final expression for the effect of
initial and final state interaction which is presented in eq.~(\ref{kappa}).
For simplicity we neglect the effects of spin rotation in the light cone spin
density function of the deuteron (see \cite{FS83}) which is a small
effect ( 1\% ) in our case.
\end{appendix}

\pagebreak
{\large {\bf Figure Captions}}
\begin{itemize}

\item[Fig. \thinspace 1.]  $F(s,\Theta_{cm})$ is the function used to best
fit the world data for $pp$ elastic scattering to the phenomenological
motiviated parametrization of \cite{RLPR} and \cite{BR:pr}. The
parametrization is ideal for large ($\geq~10~GeV^2$) $s$ and $\Theta_{cm}
\rightarrow 90^{0}$  and under this condition $F(s,\Theta_{cm}) \rightarrow 1$.

\item[Fig. \thinspace 2.] The LC PWIA cross sections for hard quasielastic
proton scattering of the deuteron as a function of the momentum ($p_s$)
and polar angle ($\theta_s$) of the spectator neutron. (a) -
$p_1~=~6~GeV/c$, (b) -  $p_1~=~12~GeV/c$.

\item[Fig. \thinspace 3.] The ratio of LC PWIA cross sections. In  the
nominator the cross section for $pp$ hard scattering was calculated with a
final state invariant energy $s$  and the denominator cross section was
calculated for the  invariant energy of the intermediate state.
The ratio is presented
as a function of the spectator momenta ($\theta_s~=~180^0$).
The dashed line is for  $p_1~=~6~GeV/c$ and  the solid line
is for $p_1~=~12~GeV/c$.

\item[Fig. \thinspace 4.] The diagrams show the PWIA and the lowest order
soft $pn$ rescattering processes. The broken line represents the soft
interaction. The empty circles represent the hard $pp$ scattering vertices.

\item[Fig. \thinspace 5.] The factor $\kappa_p$  represents
           the  effect of ISI and FSI as calculated in eikonal
           approximation according to the diagrams of Fig.4.
           The projectile momentum is $p_1~=~12~GeV/c$.
           The different lines (from top to bottom) correspond to spectator
           transverse momenta of $0,~40,~80,~120,~160~MeV/c$.

\item[Fig. \thinspace 6.] Time-ordered diagrams representing the impulse
approximation contribution (A) and the relativistic (vacuum) contribution
(B).

\item[Fig. \thinspace 7.] The ratio of the cross
	   sections as a function of $p_s$ calculated in LC over the cross
           section obtained in the VN approximation for
           $p_1~=~12~GeV/c$. (a)   parallel and antiparallel geometries,
	   (b)   perpendicular geometries (see definitions in Section 2).
	   The solid line is for the PWIA calculation with the "Paris"
           deuteron wave function.
	   The dashed line is for the PWIA calculation with the "Bonn"
           wave function. The curves with "$\Diamond$"  correspond to the
           calculation with ISI and FSI.

\item[Fig. \thinspace 8.] The $\delta (p_s)$ dependence
           of the asymmetry $A$ defined in eq.(\ref{asm}) and calculated in
           the VN approximation. $p_1~=~12~GeV/c$. The lines and symbols are
           defined in the previous figure. Note that $A~=~0$ in LC.
           The bottom scale represents the corresponding values of
           longitudinal component of the spectator momenta $p_{sz}$ calculated
           for $\alpha~=~1-\delta$ (negative $p_{sz}$) and $\alpha~=~1+\delta$
           (positive  $p_{sz}$) at $\delta = 0.1, 0.2$ and $0.3$.

\item[Fig. \thinspace 9.] The ratio of cross sections
           calculated in the LC approximation with and without taking into
           account the PLC suppression. The ratio is shown as a
           function of $p_s$ for parallel (solid), antiparallel (dashed) and
           perpendicular (dot-dashed) geometries.
           The calculations are presented for $p_1~=~12~GeV/c$ with the
            "Paris"  wave function for the deuteron.

\item[Fig. \thinspace 10.] The cross section for $p_1$ = $12~GeV/c$
            of hard quasielastic
            $pd$ scatterings as a function of spectator momentum
            at (anti)parallel geometries. Dotted line - VN PWIA,
            dash-dotted - LC PWIA,  dashed -  LC with the effect of
            PLC suppression and  solid line - LC with PLC suppression and
            with ISI, FSI. The "Paris" wave function for the
	    deuteron was used.

\end{itemize}


\begin{thebibliography}{98}

\bibitem{Brn}
	G.~E.~Brown and A.~D.~Jackson, The
Nucleon-Nucleon Interaction, \newblock { North-Holland
Publish.Comp.},1976.
\bibitem{emt}
	I.~The { et al.},
	\newblock { Phys.Rev.Lett.}, {\bf 67}, 173 (1991).
\bibitem{Arnold75}
	R.~Arnold { et al.},
	\newblock { Phys. Rev. Lett.}, {\bf 35}, 776, (1975).
\bibitem{Arnold87}
	R.~Arnold { et al.},
	\newblock { Phys. Rev. Lett.}, {\bf 58}, 1723, (1987).
\bibitem{Rock}
	S.~Rock { et al.},
	\newblock { Phys. Rev. Lett.}, {\bf 49}, 1139, (1982).
\bibitem{pd:el}
	G.~Alberi and G.~Goggi,
	\newblock { Phys. Rep.}, {\bf 74}, 1, (1981).
\bibitem{fb1}
	A.~M.~Baldin et al., { preprint JINR}, Dubna, (1977).
\bibitem{fb2}
	L.~M.~Anderson Jr., { preprint LBL}, 6769, Berkley, (1979).
\bibitem{FS:rep81}
	L~.L.~Frankfurt and M.~I.~Strikman, { Phys.Rep.},
{\bf 76}, 214, (1981).
\bibitem{FS:rep88}
	L.~L.~Frankfurt and M.~I.~Strikman, { Phys.Rep.},
{\bf 160}, 235 (1988).
\bibitem{FS79}
	L.~L.~Frankfurt and M.~I.~Strikman, { Nucl.Phys.},
{\bf B148}, 107, (1979).

\bibitem{Eva}
	A.~S.~Caroll et al., { BNL Experiment}, 850.
\bibitem{Cebaf}
	Conceptual Design Report, CEBAF 1990.
\bibitem{FFS}
	G.~R.~Farrar, L.~L.~Frankfurt and M.~I.~Strikman,
	{ Phys. Rev. Lett.}, {\bf 62}, 1095, (1989).
\bibitem{FS85}
	L.~L.~Frankfurt and M.~I.~Strikman, { Nucl.
Phys.}, {\bf 250}, 143  (1985).
\bibitem{RLPR}
	J.~P.~Ralston and B.~Pire, { Phys.Rev.Lett.},
        {\bf 61}, 1823,  (1988).
\bibitem{BR:pr}
	S.~J.~Brodsky, High Energy Collisions - 1973
(Stony Brook) Conf.Proc.
\bibitem{ARNDT}
	R.~A.~Arndt,
	{ Phys.ReV.} {\bf D 35}, 212, (1987).
\bibitem{FS83}
	L.~L.~Frankfurt and M.~I.~Strikman, { Nucl.
Phys.}, {\bf A405}, 557, (1983).
\bibitem{West}
	W.~B.~Atwood and G.~B.~West, { Phys. Rev.},
{\bf D 7}, 773, (1973).
\bibitem{nm}
	L.~L.~Frankfurt and M.~I.~Strikman,
{ Phys.Lett.}, {\bf 64B}, 435, (1976).
\bibitem{KOSP}
	J.~B.~Kogut and D.~E.~~Soper, {
Phys.Rev.}, {\bf D 1}, 2901, (1970).
\bibitem{Bonn} R.~Machleidt, K.~Holinde and C.~Elster,
	{Phys.Rep.}, {\bf 149}, 1, (1987).
\bibitem{Paris} M.~Lacombe et al.,
	{ Phys.Rev.}, {\bf C 21}, 861, (1980).
\bibitem{FMS}
	L.~L.~Frankfurt, G.~Miller and M.~I.~Strikman,
	{ Phys.Rev.Lett.} {\bf 68}, 17, (1992).
\bibitem{Yuan}
	Jin - Yuan Wu, Thesis,
	{ Pennsylvania State University } 1992.

\bibitem{FS91}
	L.~L.~Frankfurt and M.~I.~Strikman, {
Progress in Part. and
	Nucl. Phys.} {\bf 27}, 1991.

\bibitem{KAF}
	G.~K\"{a}lbermann, L.~L.~Frankfurt and J.~M.~Eisenberg,
{ Submited to Phys.Lett.B } (1994).

\bibitem{ptf}
	L.~L.~Frankfurt, M.~I.~Strikman and M.~Zhalov,
to be published.

\end{thebibliography}
\end{document}